\documentclass{article}
\usepackage[utf8]{inputenc}
\emergencystretch=\maxdimen
\usepackage[utf8]{inputenc}
\usepackage[hidelinks]{hyperref}
\usepackage{xcolor}
\usepackage{soul}
\usepackage{graphicx}
\usepackage{bm}
\usepackage{float}
\usepackage{mdframed}
\usepackage{algorithm}
\usepackage{algorithmicx}
\usepackage{algpseudocode}
\usepackage[english]{babel}
\newtheorem{risk}{Risk}
\newtheorem{mitigation}{Mitigation}
\newtheorem{target}{Target}
\newtheorem{definition}{Definition}
\newtheorem{incentives}{Incentive}

\usepackage{amsmath}
\usepackage{authblk}
\usepackage{amssymb}
\usepackage{longtable}
\emergencystretch=\maxdimen
\usepackage{geometry}
\usepackage{lipsum}
\newcommand\blfootnote[1]
{
  \begingroup
  \renewcommand\thefootnote{}\footnote{#1}
  \addtocounter{footnote}{-1}
  \endgroup
}

\title{Proof of Efficient Liquidity:\\A Staking Mechanism for Capital Efficient Liquidity}
\author{Arman Abgaryan*$^{I, II}$, Utkarsh Sharma*$^{I, III, IV}$, Joshua Tobkin$^I$}
\date{October 2023}

\begin{document}
\maketitle

\blfootnote{*Lead co-authors.}
\blfootnote{I. Supra}
\blfootnote{II. Shanghai Jiao Tong University}
\blfootnote{III. Oxford-Man Institute of Quantitative Finance}
\blfootnote{IV. University of Oxford}

\begin{abstract}
    The Proof of Efficient Liquidity (PoEL) protocol, designed for specialised Proof of Stake (PoS) consensus-based blockchains that incorporate intrinsic DeFi applications, aims to support sustainable liquidity bootstrapping and network security. This concept seeks to efficiently utilise budgeted staking rewards to attract and sustain liquidity through a risk-structuring engine and incentive allocation strategy, both of which are designed to maximise capital efficiency. The proposed protocol serves the dual objective of: (i) capital creation by attracting risk capital efficiently and maximising its operational utility for intrinsic DeFi applications, thereby asserting sustainability; and (ii) enhancing the adopting blockchain network’s economic security by augmenting their staking (PoS) mechanism with a harmonious layer seeking to attract a diversity of digital assets. Finally, the protocol’s conceptual framework, as detailed in the appendix, is extended to encompass service fee credits. This extension capitalises on the network’s auxiliary services to disperse incentives and attract liquidity, ensuring the network achieves and maintains the critical usage threshold essential for its sustained operational viability and progressive growth.
\end{abstract}

\section{Introduction}
    A blockchain is a specialised immutable database operating in a decentralised manner utilising distributed computing resources, employing consensus algorithms, and acting as a secure technological platform to facilitate the execution of transactions. These platforms evolve into ecosystems with digital assets facilitating the exchange of value, assuring economic security, and enabling work. Specifically, note that some of these blockchain networks have core decentralised finance (DeFi) applications (CDAs), which are specific financial applications that are infrastructurally imperative and intrinsically integrated into a blockchain network and its economic framework, enabling the network to meet its objectives. These applications operate using pools of digital assets submitted by liquidity providers (LPs) in smart contracts to facilitate financial transactions, like lending, market making, etc. Naturally, such applications often face a paradoxical causal loop, where they need a minimum viable capital to be operational and attract users, which can only happen if the application experiences sufficient demand to offer compelling returns. Furthermore, the blockchain network underlying the CDAs runs in a decentralised and distributed manner, necessitating the use of consensus mechanisms which enable the network to agree, confirm, and validate transactions. One example of such a consensus mechanism is the Proof-of-Stake (PoS) mechanism where locked (staked) capital (in the form of digital assets) is used to secure the network and assure operational integrity. Therefore, for continuing success of not just CDAs, but also the underlying blockchain network using the PoS consensus mechanism is directly linked to: (a) the amount of (locked) capital it can attract from participants; and (b) the incentives and penalties it uses to encourage and/or discourage behaviours on the network, to harmonise experience.\\
    \\ 
    Whilst blockchain networks operating in a Proof of Stake (PoS) consensus mechanism often seek to attract staked capital in their native staking token — a digital asset that users can lock or \lq \lq stake\rq\rq to participate in network operations like transaction validation, in return for potential rewards, and thereby help maintain the economic security and functionality of the blockchain, with stakers often being rewarded for their contribution — as their ability to sustainably attract new capital might be limited, particularly if the perceived risk-adjusted return is not immediately compelling. This means that competition for the acquisition of capital from stakers (including node operators), whose capital base may already be predisposed, leads to an interdependent quandary. Staked capital is only present when the risk-adjusted return is compelling, but this is hard to sustainably achieve until the network has baseline-staked seed capital (economic security guarantees) to be an attractive platform for its users. This creates uncertainty for the network in ensuring the participation of a critical number of service providers and staked capital for operational integrity. A similar quandary exists for participants who have access to capital, but are averse to a direct exposure to the native staking token of the network. This is partly because in the early stages of the network, the price of its native staking token may be volatile\cite{pessa2023agemc}, and it leads to fluctuation in PoS security guarantees of a network associated with the staked asset's value. Therefore, constrained diversification of the staked asset base has a disproportionately desirable potential to improve the robustness of security guarantees of the network.\\
    \\
    We seek to resolve this paradoxical causal loop through a budgeted incentive program wherein PoEL, an algorithmic program designed to support blockchain networks with CDAs, effectively draws operating capital and enhances the economic realities of the underlying network, encompassing ongoing viability and economic security. The proposed incentive framework facilitates a controlled expansion of the utility of capital submitted to CDA, making it available as staked capital for the network integral to CDAs, thus providing an additional income stream for capital providers. This is achieved by offering risk-aware preferential incentives over and above regular (CDA) fees, which may be denominated in the native staking token of the adopting blockchain network, or otherwise. These incentives allow digital asset holders to fully utilise their risk capital for on-chain economic activities. Such on-chain economic activities can initiate a self-reinforcing cycle where a blockchain network's core DeFi application seeks to attract risk capital for on-chain economic work, which also secures the network. As the capital base improves, it stabilises the economic fundamentals, which in turn attracts more capital. However, the extent to which CDA capital is available for staking, and its resulting eligibility for additional earnings, is governed by the PoEL protocol.\\
    \\
    PoEL achieves the aforementioned vision through a two-pronged approach. First, it solves the interdependent quandary of the minimum viable amount of capital required by the CDAs and the underlying blockchain infrastructure, by serving as a novel mechanism for preferential incentives to attract liquidity. Second, it enables networks to efficiently attract capital locked in a diversity of LP pools through its novel interest rate and collateral guidance framework enabling the risk-conscious extension of the utility of the attracted capital. This has the effect of optimising a diversity of risks and rewards to effectively integrate CDA liquidity pools with the network's own PoS scheme, thereby, improving the native network's economic security. It is noteworthy that the proposed framework does not seek to propose an alternative PoS mechanism but rather design an auxiliary protocol which works in conjunction with existing PoS mechanisms for networks with integrated CDAs, to optimise their incentive programs to attract capital. This general framework enables digital asset owners to seek excess returns on their locked capital and enables CDAs to bootstrap liquidity, and networks to mitigate asset-specific risks, thereby improving risk-adjusted returns generated per unit of network resources.\\
    \\
    The remainder of the paper is organised as follows: in Sec. \ref{sec:litreview} we discuss benchmark works of relevance; in Sec. \ref{sec:principles} we outline key principles governing the protocol's architecture; in Sec. \ref{sec:protocol} we describe the protocol's architecture, before concluding in Sec. \ref{sec:conclusion}.

\section{Literature Review}\label{sec:litreview}
    Digital asset-based liquidity bootstrapping in decentralised finance, also referred to as \lq\lq liquidity mining,\rq\rq was first introduced by a decentralised exchange IDEX \cite{idex}, and thereafter improvised by their competitors, e.g. Synthetix \cite{synthetix} and Compound \cite{compund}. Such developments paved the way for a period of significant growth for the digital assets industry, reflected in a sharp increase in Total Value Locked (TVL) from \$600 Million to \$11 Billion between March and September of 2020 \cite{cousaert2022sok}. The approach to use additional rewards (denominated in native tokens) to attract liquidity has been adopted by different types of DeFi projects, like money market applications, for e.g., Compound, AAVE \cite{aave}, Benqi \cite{benqi}, Alchemix \cite{alchemix}; AMM/DEX protocols, such as Uniswap \cite{uniswap}, Sushiswap \cite{sushiswap}, Curve \cite{curve}, Balancer \cite{balancer}; insurance protocols like Nexus Mutual \cite{nexus_mutual}, InsurAce \cite{insurace}; and derivative platforms like Perpetual Protocol \cite{prepetual}, dYdX \cite{dydx}, Synthetix \cite{synthetix}. Some of these projects, use governance tokens \cite{fan2023towards} as a mechanism of rewarding additional liquidity. For context, methods of bootstrapping liquidity that are now popular in this domain can trace their origins in traditional finance (e.g. see: \cite{clapham2021liquidity, dosanjh2013market, menkveld2013designated}).\\
    \\
    Related to the concept of liquidity bootstrapping in the digital finance domain, some works that might be helpful for context include the design for UniSwap's V3 \cite{Yinh2021liqmining} and digital asset exchanges, e.g. \cite{feng2019liquidity}. Broadly speaking, liquidity bootstrapping strategies implement reward mechanisms to incentivise liquidity, which could be quantified based on different factors. E.g. Curve \cite{lgcurve} used a decentralised governance mechanism to quantify rewards using factors like asset pool weights and characteristics, and similarly, Balancer \cite{lmbalancer} quantified rewards based on factors like the type of asset, size of the pool and its state, and pool-specific fees. Once calibrated, these rewards are distributed either primarily to LPs (like in the case of Uniswap, Aave, and Curve Finance), or can be extended beyond LPs to stakeholders like traders and borrowers (like in the case of Osmosis, and Compound). However, some of the liquidity mining programs not only incur direct costs but may also expose the adopting applications to indirect costs, stemming from risks leading to manipulative abuse of promotional offerings. Further, the industry has also experienced noteworthy risk events, e.g. wash trading(and by extension, wash transactions involving credit, as witnessed by Compound) like manipulative tactics, as witnessed by FCoin's \cite{fcoincoindesk} liquidity mining program, and flooding of opportunistic hot money - seeking to solely pocket any incentives and dump the digital asset once incentives are collected, instead of using it as an opportunity to be a meaningful participant in an ecosystem.\\
    \\
    Therefore, our research seeks to present a cost and risk-conscious liquidity bootstrapping program, to enable blockchain networks with integrated core DeFi applications (CDAs), to maximise the utility of every unit of risked capital, whilst harmonising economic interests of the underlying network and the implementing CDAs. A few examples of networks that have integrated CDAs for cross-chain asset exchange and derivative use-cases include Osmosis \cite{osmosis}, and Qasar \cite{qasar}. The proposed bootstrapping mechanism is most closely related to Osmosis' \cite{ssosmosis} incentive mechanism, which not only rewards LPs using AMM fees, but additionally rewards them for value created from assets staked by them in the network, i.e. staking rewards in exchange for assets provided to secure the network. However, the proposed protocol design is distinct from their approaches, as PoEL jointly focuses on capital efficiency and network security, by accommodating a set of well-defined principles and objectives.

\section{Objectives}\label{sec:principles}
    PoEL is defined as an algorithmic policy function, $\mathbf{\Pi}(\mathcal{M}(\kappa, R)): \mathbf{S}_t \rightarrow \bm{\mathcal{I}}_{t+1}$, which maps state observables ($\mathbf{S}_t$) to a vector of incentives ($\bm{\mathcal{I}}_{t+1}$), by using its liquidity mining program ($\mathcal{M}(\kappa, R)$), to essentially solve an optimisation problem constrained on the chosen risk measure $\kappa$ and total distributable incentive budget ($R$). Whilst being backwards compatible with many existing blockchain networks' PoS schemes, this design ensures efficient preservation of any financial value being created for the network, by aiming to maximise the expected financial value of total transactional volume and minimise the variance of the value of the asset mix backing the security of the network, subject to stochastic constraints.\\
    \\
    Before we progress, we seek to specify a few important concepts:

    \begin{definition}[Timestep]
        A timestep ($t$) refers to a specific point in time when a change in state variables is observed, or input parameters are set.
    \end{definition}
    
    \begin{definition}[Epoch]
        An epoch ($e$) is a short, yet relatively fixed period, akin to block time, representing instances when the protocol executes critical updates. These updates include, but are not limited to, instances such as reward distribution, accepting new users to the system, and enforcement of penalties on participants.
    \end{definition}

    \noindent
    It is important to note that whenever a measure, say - price, carries a subscript $e$, i.e. it is the representative price of the epoch $P_e$, it means that this is the price available in the last timestep of the epoch, i.e. $P_e = P_{T}, \forall e:[0,T]$.

    \begin{definition}[Unstaking Period]
        The unstaking period refers to the predetermined duration of time set by a blockchain network or DeFi platform, which follows a request to withdraw (or unstake) staked assets when the digital assets remain inaccessible or non-transferable.
    \end{definition}

    \noindent
    The unstaking period is designed to promote a blockchain network's security and stability, and to discourage malicious activities, by preventing abrupt changes in the staked capital and limiting the ease of materialising a malicious act.\\
    \\
    Now, we seek to optimise over the decision variable $\bm{\mathcal{I}}$, which is the incentive allocation matrix capturing available incentives across asset pools of a particular CDA and epochs, and present the forthcoming optimisation statement as a framework to motivate the conceptual design of the PoEL protocol presented in sections that follow.
    
    \begin{equation}\label{eq:mainopt}
        \begin{aligned}
            & \underset{\bm{\mathcal{I}}}{\text{Maximise}} \quad w^\prime_1 \cdot \sum^{\epsilon}_{q=1} \mathbb{E}[V^i_q] - w^\prime_2 \cdot \sum^{\epsilon}_{q=1}  \mathbb{E} [\sigma^2_q]
            & \textbf{s.t.}\\
            & \mathbb{P}\left(\sum^l_{o=1} \bm{\kappa}^{o}_{e} \leq \bm{\kappa}^i_e\right) \geq \alpha\\
            & cVaR(\bm{\mathcal{I}}_e)_{\beta=0.99} \leq Limit\\
            & \sum^\epsilon_{q=1} \sum^l_{o=1} \mathcal{I}_o^l = \text{R}
        \end{aligned}
    \end{equation}

    \noindent
    where, $w^\prime_1$ and $w^\prime_2$ are the weights representing the importance of each objective, s.t. $w^\prime_1 + w^\prime_2 = 1$; $\epsilon$ is the length of incentivisation program in epochs;  $\mathbb{E}[V_q^i]$ is the expected financial value for the network of total transactional volume in $q$-th epoch; $\bm{\mathcal{I}}_q$ is vector of incentives distributed across asset pools in $q$-th epoch, representing one particular column in the matrix $\bm{\mathcal{I}}$; $\mathcal{I}^o_e$ represents incentives distributed for asset pool $o$ in e-th epoch; $l$ is the number of asset pools of CDA $i$, included in the PoEL program; $\mathbb{P}\left(\sum^l_{o=1} \bm{\kappa}^{o}_{e} \leq \bm{\kappa}^i_e\right) \geq \alpha$ is a probabilistic constraint ensuring that the aggregate risks of the portfolio of asset securing the network do not exceed a certain limit ($\kappa^i_e$) with a probability of at least $\alpha$; $cVaR(\bm{\mathcal{I}}_e)_{\beta=0.99}$ ensures that the  Conditional Value at Risk (cVaR), calculated at a pre-specified confidence interval ($\beta$) stays below a pre-specified limit\footnote{Note that the protocol aims to limit the expected shortfall, as indicated in the constraint. It seeks to quantify this by using the portfolio returns of assets. These assets are attracted using the epoch-specific incentives, which is the decision variable in the objective statement of the preceding optimisation problem.}; and $\mathbb{E} [\sigma^2_q]$ is the expectation of the variance of the marked-to-market value of the assets backing the economic security of the network\footnote{We express the financial value of transactional volume and the variance in nominal terms.}.\\
    \\
    This representative optimisation problem is tackled using the proposed protocol, by way of enabling varied market forces to discover points of dynamic equilibrium (i.e. either local or global optimal solution), reflecting the protocol's goal of accommodating the requirements and strategies of each rational economic actor. Our deployment of this optimisation framework serves to establish the context for the operating and financial principles that follow, along with the associated objectives, as detailed in the protocol architecture presented in Section \ref{sec:protocol}.

\subsection{Operating Principles}

    \begin{risk}[Wash Transactions]\label{risk-washtransactions}
        A wash transaction is defined as the predatory activity of transacting (e.g. buying, selling, lending, or borrowing) in an asset, with the sole objective of creating the false impression of market activity, without any meaningful change in ownership, or financial impact.\\
        \\
        Using the example of a decentralised exchange (DEX), if $P_t$ is the price of the asset, and $\Delta V_t$ is the volume traded, we can state that a trade is a wash trade if, for e.g., the DEX experiences the following sequence of trades:

        \begin{align*}
            \text{Buy(t)}: & \quad P_t \cdot \Delta V_t > 0\\
            \text{Sell(t+1):} & \quad P_{t+1} \cdot \Delta V_{t+1} < 0,\\
            \text{Trade:} & \quad P_t - P_{t+1} \sim 0 \quad \text{\&} \quad |\Delta V_t| = |\Delta V_{t+1}| > 0
        \end{align*}
    \end{risk}

    \begin{mitigation}[Wash Transactions]\label{mitigate-washtransactions}
        We seek to discourage wash transactions, using the following tools: 
        \begin{enumerate}
            \item Mitigation: Disincentivising wash transactions, by e.g. financially incentivising organic user growth, by ensuring that the cost exceeds any expected gains from sustained malicious attempts.
            \item Transformations: Transforming point-in-time measures used to drive reward distribution policy to moving averages, makes it harder for malicious actors to manipulate key metrics through wash transactions.
        \end{enumerate}
    \end{mitigation}

    \begin{risk}[Continuous Quality Assurance]\label{risk-quality}
        In a PoS mechanism, there is a risk that in the event of an adverse change in the fair value of the staked asset, the network's economic security might be compromised, if assets backing the network security are not marked-to-market.
    \end{risk}

    \begin{mitigation}[Continuous Quality Assurance]\label{mitigate-quality}
        The system continuously checks the market price of assets used to secure the network, and exhibits demonstrable intention to make adjustments to assertively tackle any accounting discrepancies which emerge.
    \end{mitigation}

    \begin{risk}[Network Liveness]\label{risk-liveness}
        PoEL's operational integrity principle focuses on validator sustenance\footnote{Validator sustenance refers to a validator's ability to actively and effectively perform its role in the adopting network.} ($OL_1$) where the protocol seeks to maximally avoid all scenarios which increase, or even reinforce the likelihood of a cascading number of nodes from being disqualified from a PoS-based blockchain network, as a result of failing to meet the dynamic financial requirements, e.g. minimum staking requirements \footnote{PoS-based blockchain networks often impose a minimum staking requirement to ensure that participants have a vested interest in the network's stability and security. By requiring a minimum amount of tokens or cryptocurrency to be staked, the network can discourage frivolous or malicious actors who might otherwise participate without a significant investment, by slashing their staked capital. This requirement helps to align the interests of the stakeholders with the long-term health and integrity of the blockchain network.}.
    \end{risk}

    \begin{mitigation}[Network Liveness]\label{mitigate-liveness}
        Operational breaches are mitigated by maintaining a breach-specific threshold, such that the event-specific probability is managed as follows: 
        \begin{equation}
            \mathbf{P}[OL_1] \leq \lambda_{{OL}_1},
        \end{equation}
        
        \noindent
        where $\lambda_{{OL}_1}$ is the probabilistic threshold for the likelihood of the occurrence of prespecified operational risk events. And since $\lambda_{{OL}_1} \neq 0$, in its simplest form we seek to achieve this by incentivising (or imposing) only limited and operationally robust validators to participate in PoEL, thereby reducing the risk of cascading disqualifications which would fundamentally compromise the operational integrity of the network.
    \end{mitigation}

    \begin{risk}[Backward Compatibility]
        If the protocol is not backwards compatible with the adopting network's staking mechanism\footnote{The PoEL protocol specifically seeks to be compatible with some existing networks, which use smart contracts to manage staking aspects of their system.}, it would limit the PoEL protocol's real-world applicability. This limitation could impact its viability.
    \end{risk}      

    \begin{mitigation}[Backward Compatibility]
        The proposed protocol should not require any major changes to the adopting network's staking dynamics.
    \end{mitigation}

    \begin{risk}[Stake Centralisation Attack]\label{risk-majority}
        A stake centralisation attack refers to an event where a user (or a group of users) amasses such a significant portion of the total staked capital in the PoS-based network, that it undermines the security of a decentralised network as the user(s) effectively control over the network's decision-making, block creation and transaction validation processes.
    \end{risk}

    \begin{mitigation}[Stake Centralisation Attack]\label{mitigate-majority}
        The generally accepted shape of the liquidity density function of any financial asset, whether digital or traditional, implies that the adverseness of execution price (or the resulting market impact) from the execution of trade is positively correlated to the executed volume, and negatively correlated to the liquidity of the asset. On a ceteris paribus basis, if a single token is used for staking in a blockchain network, a malicious stakeholder seeking to attempt to buy a significant portion of the native staking token\footnote{A native staking token is a digital asset used for staking in its associated blockchain network, securing the network and enabling stakers to earn rewards in exchange for assuring the quality of the services provided by the network. Native staking tokens (NST) are key elements of a Proof of Stake (PoS) consensus mechanism.} will have to incrementally pay a higher cost of execution, by way of higher market impact (especially if the adopting network is in its infancy and the native staking token has limited liquidity), compared with having the capacity to submit multiple assets to satisfy their desired staking outcome, i.e.
        \begin{equation}
            MI^{\phi_{NST},V}_{NST} \geq \sum_{j=1}^N MI^{{\phi_j}, V_j}_j,
        \end{equation}

        \noindent
        where $MI^{\phi_{NST},V}_{NST}$ represents the market impact of executing a fixed notional value $V$ in the native staking token with a market impact function $\phi_{NST}$, and $MI^{{\phi_j}, V_j}_{j}$ represents the market impact of executing a smaller notional value $V_j$ across N-many assets with their respective market impact functions $\phi_j$, such that $\phi_1 = \phi_2 = \dots = \phi_N = \phi_{NST}$.\\
        \\
        However, since market impact functions ($\phi$) are not all the same, a situation arises where the superiority of a market impact function, i.e. its ability to absorb large volumes, is directly related to the ease with which a malicious stakeholder can attempt (or successfully execute) a stake centralisation attack.\\
        \\
        Therefore, to ensure that the incentive mechanism that enables the use of multiple assets as staked capital to secure an early-stage network, does not end up easing natural market-based barriers to a stake centralisation attack, we seek to limit the portion of capital staked through other digital assets which aren't a native staking token, as follows: 
        
        \begin{equation}
            \frac{V_{MA}}{V_{MA} + V_{NST}} < T
        \end{equation}
        where $V_{MA}$ is the value staked in multiple assets, and not the native staking token, $V_{NST}$ value staked in the network in native staking token and $T$ is the optimal threshold which mitigates the risk. 
        
    \end{mitigation}

    \begin{risk}[Incentive Disharmony]\label{risk-disharmony}
        Incentives provided as part of the PoEL protocol may have a predatorial effect on the incentive structure of the adopting network\footnote{Absence of incentive harmony can, e.g., distort a capital provider's motivation to select the best validators, leading to an important agent like validator being delegated-to without merit.}. 
    \end{risk}

    \begin{mitigation}[Incentive Disharmony]\label{mitigate-disharmony}
        To mitigate the risk of PoEL's incentive structure being disharmonious to the incentive structure of the adopting blockchain network, the PoEL protocol penalises destructive (malicious and/or non-contributory) stakeholders by slashing their locked capital.
    \end{mitigation}
    
    \noindent
    This framework and foreseen benefits ought to be balanced against additional risks it poses, which if unchecked, leaves the entire system more vulnerable in times of extreme volatility. Therefore, we seek to ensure that the PoEL protocol considers a variety of risks, including, but not limited to market risk.
    
\subsection{Financial Principles}
    \begin{risk}[Realised Volatility]\label{risk-vol}
        Variability, for e.g., as measured by volatility, in the basket of assets backing the network introduces stochasticity in its economic security. This can be quantified at an asset level, as follows:
        \begin{equation}
            \sigma = \sqrt{\frac{1}{N-1} \sum_{t=1}^{N} (r_t - \bar{r})^2}
        \end{equation}
        \noindent
        where $r_t$ represents an asset's return, and $N$ is the number of assets in the portfolio.
    \end{risk}
    
    \begin{mitigation}[Realised Volatility]\label{mitigate-vol}
        The system homogenises the risk-adjusted desirability of individual assets and enforces the volatility requirements on the portfolio of assets utilised in PoEL.
    \end{mitigation}

    \begin{risk}[Tail Risk]\label{risk-tail}
        In addition to the preceding risk which focuses on mitigating risks emerging from the second moment of returns distribution, which can be somewhat said to be \lq\lq ongoing\rq\rq risks, the network's economic security is also exposed to unexpected losses, for e.g. as captured by measures like the Expected Shortfall\cite{acerbi2002coherence}, which may also additionally exacerbate the risk of cascading disqualification of validators.
    \end{risk}

     \begin{mitigation}[Tail Risk]\label{mitigate-tail}
        To mitigate the tail risk, the protocol's objective statement is constrained by an expected shortfall limit.
    \end{mitigation}

    \begin{risk}[Liquidity Risk (Viscosity)]\label{risk-liquidity}
        Liquidity risk is the risk that the protocol would be unable to buy or sell an asset without leading to a meaningfully adverse price impact, where the adverse price impact is quantified as $LR$, and in a simplified form, it can be expressed as a function of trading volume, bid-ask spread, and order size.
        \begin{equation}
            LR = f(Sp,V^{t_1:t_2},Q),
        \end{equation}
        \noindent
        where $V^{t_1:t_2}\in [0, \infty)$ is the trading volume of the asset over a lookback window, $Sp \in \mathop{R}^+$ is the bid-ask spread, and $Q$ is the order size.
    \end{risk}

    \begin{mitigation}[Liquidity Risk (Viscosity)]\label{mitigate-liquidity}
        A minimum liquidity requirement is asserted for each of the qualifying assets.
    \end{mitigation}

\subsection{Ancillary Objectives}
    Based on the preceding principles, we now outline the ancillary objectives PoEL protocol seeks to target, and the incentives used to catalyse their realisation. 
    
    \begin{target}[Tenure]
        To incentivise the stickiness of capital submitted to liquidity pools and its subsequent use for network security, attributable incentives are positively correlated to the length of time capital is locked. 
    \end{target}

    \begin{incentives}[Tenure]
        \begin{equation}
            \mathcal{I}_t^l = \mathcal{I}_{t}^{l, {\max}} + \frac{\mathcal{I}_t^{l, {\max}} - \mathcal{I}_t^{l, \min}}{(1 + e^{-k (t^l - t^l_{\text{mid}})})^\varrho}
        \end{equation}

        \noindent
        where, $t^l$ represents the time for which $l$-th LP capital is locked, $\mathcal{I}_t^l$ represents the associated incentive, $\mathcal{I}_t^{\max}$ represents the maximum incentive available to distribute to a particular LP at a particular timestep, $\mathcal{I}_t^{l,\min}$ is the minimum incentive (portion of $\mathcal{I}^{l, {\max}}_t$), $k$ is a scaling factor that adjusts the steepness of the sigmoid curve, $t_{\text{mid}}^l$ is a mid-point value around which the transition from the minimum to maximum incentive starts to become significant, and $\varrho$ is a new parameter that adjusts the curve's shape, providing additional control over how rapidly the incentive increases as locking time approach and surpasses mid-point.
        
        \begin{figure}[H]
        \begin{center}
            \includegraphics[scale=0.5]{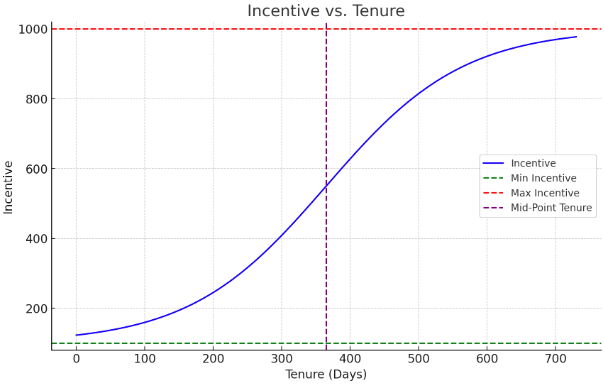}
            \caption{Sliding scale of tenure-based incentives, with $\mathcal{I}_{\min} = 100$, $\mathcal{I}_{\max} = 1000$, $t_{\text{mid}}^l = 365$, $k = 1$, and $\varrho = 0.01$.}
            \label{fig:slidingtenure}    
        \end{center}
        \end{figure}

        \noindent
        Now, the role of an investor seeking to lock in an amount would be to quantify the present value of these forward incentive rates at a point in time $t$, as follows:
        \begin{equation}
            PV^l = \int_{t_L=0}^{t_L=t} \mathcal{I}^l_t(t_L).e^{-\int_{u=0}^{u={t_L}}IR(u)du}dt_L,
        \end{equation}
    
        \noindent
        where the integral captures the continuously compounding forward incentive rates\footnote{We assume that continuous compounding would be harmonious with the adopting blockchain network's economics.}, by additionally introducing an interest rate ($IR(t)$) on top of the incentive rate, aimed at capturing the slope of the interest rate curve. Note, that $t_L$ is a variable of integration, representing the time when incentive's received.
    \end{incentives}

    \begin{target}[Capital Productivity ]\label{target-ce}
        Capital productivity is defined as the productive utilisation of the attracted capital, where productivity is defined as a net positive utility of the expected return (incl. any additional fees generated through increased volume) when adjusted for the cost of capital.
    \end{target}

    \begin{incentives}[Capital Productivity]\label{incentive-ce}
        We assume market participants to be rational, and therefore, the PoEL protocol assumes that given its sustainable and competitive structure which dynamically responds to changes in exogenous factors, capital productivity is either always achieved, or incentive structures are updated such that there is progress in its direction.
    \end{incentives}
    
    \begin{target}[Responsiveness]\label{target-responsiveness}
        The protocol aims to be responsive to changes in state variables($\mathbf{S} = \{S_0, S_1,\dots, S_i\}$), to incentivise actions which seek to progress the succeeding state towards its stated goals.
        \begin{equation}
            \frac{\partial \bm{\mathcal{I}}}{\partial \mathbf{S}} \neq 0 
        \end{equation}
    \end{target}
    
    \begin{incentives}[Responsiveness]\label{incentive-responsiveness}
        To achieve the aforementioned condition, it is designed such that it necessitates that any change in incentives progresses the protocol towards its goals, i.e.:
        \begin{equation}
            \Delta \mathcal{I} \rightarrow \text{Minimise} (|\text{State}_t - \text{Objective}^\ast|) 
        \end{equation}
    \end{incentives}

\section{Protocol}\label{sec:protocol}
\begin{figure}[H]
    \begin{center}
        \includegraphics[scale=0.6]{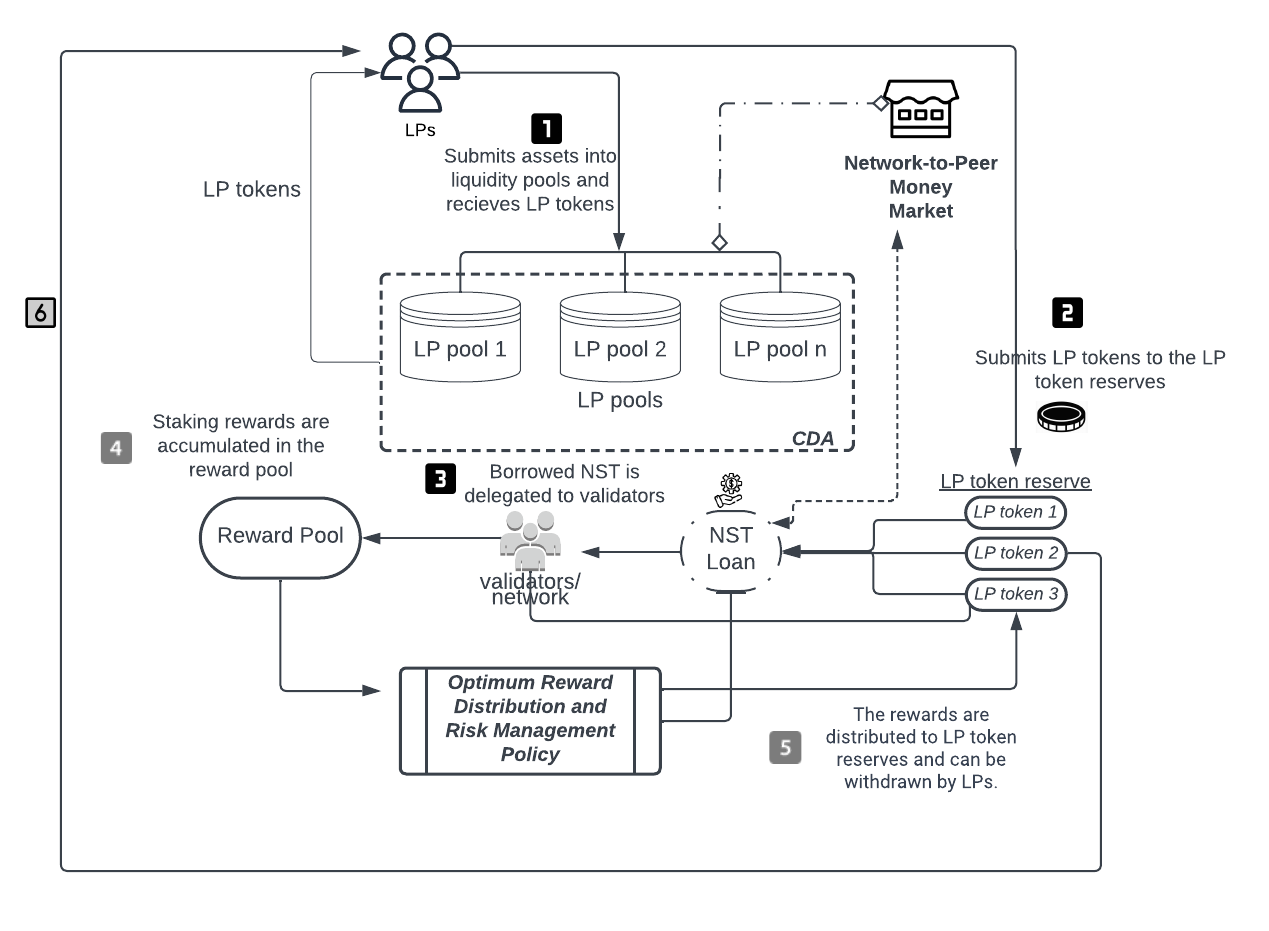}
        \caption{Schematics of PoEL}
        \label{fig:helicopterview}    
    \end{center}
    \end{figure}

    The preceding sections highlighted that PoEL seeks to aid in sustainably attracting capital to the liquidity pools of a CDA, while efficiently asserting the underlying network's economic security. This can be understood as optimising the multiplier of a CDA's operating capital available as staked capital for the underlying network. As such, PoEL is tantamount to a money market protocol in which the lender is the blockchain network, and the borrower is loaned native staking tokens in exchange for multi-asset collateral, without the associated credit risk. This is achieved through the architecture visualised in the schematics presented in Fig. \ref{fig:helicopterview}.
    
    \noindent
    We now proceed to discuss the operational mechanics of this architecture: when an LP submits asset $X$ to the i-th CDA's liquidity pool, they receive LP tokens denoted by $\tilde{X}^i$, representing an LP's share in an asset-specific pool, which can be used to further borrow native staking tokens from the network by depositing assets (LP tokens) into the LP token reserve \footnote{If a CDA's design does not assume distribution of LP tokens to its capital contributors, other approaches can be adopted to record the capital contributors willingness to use its liquidity as collateral to participate in the incentive program}. 
    
    \begin{definition}[LP Token Reserve]
        An LP's token reserve of size - $S^{\tilde{X}^i}_{e_v}$, in a particular epoch, is a unique collateral pool, which is mapped to a particular validator's address (say, v-th validator), and $\tilde{X}^i$ represents the type of asset (LP token) which can be deposited in the specified reserve.
    \end{definition}
    
    \noindent
    We denote the sum of the size of all LP token reserves including asset $\tilde{X}^i$ mapped to all validators by $S^{\tilde{X}^i}_{e} \in \mathop{\mathbb{R}}$, $S^{\tilde{X}^i}_{e} = \sum_v S^{ \tilde{X}^i}_{e_v} $, where $v$ represents the number of validators involved in the PoEL protocol. Once LP submits collateral (LP tokens) to a reserve pool, and receives a loan of the native staking token (where the amount depends on collateralisation rate ($\rho_e^{\tilde{X}^i} \in [1, \infty)$)), it is delegated to a particular validator ($v$). Notably, the amount of the loan extended to the LP is continuously tested against dynamically changing collateralisation requirements, which have to be met at all times, by all borrowers, i.e. for the total amount of native staking tokens ($L_e \in [0, \infty)$) loaned to all borrowers.
    
    \begin{definition} [Collateralisation Rate ($\rho_e^{\tilde{X}^i}$)]
        The collateralisation rate quantifies the amount of collateral an LP needs to deposit, to be able to borrow desired native staking tokens. It is represented by  ($\rho_e^{\tilde{X}^i}$), for LP token $\tilde{X}^i$, and calculated as follows:
        \begin{equation}
             \rho^{\tilde{X}^i}_e = \frac{(S^{ \tilde{X}^i}_{e}) \cdot P^{\tilde{X}^i}_e}{L^{\tilde{X}^i}_e},
        \end{equation}
        
        \noindent
        where $P^{\tilde{X}^i}_e \in \mathop{\mathbb{R}}^+$ is the price of LP asset $\tilde{X}^i$ denominated in the native staking token at the epoch $e$, and $L^{\tilde{X}^i}_e \in \mathop{\mathbb{R}}^+$ represents the loaned amount of the native staking token, using collateral in asset $\tilde{X}^i$. 
    \end{definition}
    
    \noindent
    For securing the network with their borrowed native staking tokens, LPs can earn epoch-specific staking rewards from the network. Once accrued, staking rewards are deposited in the reward pool, which represents the total budget available for the liquidity bootstrapping program, and $RP_e \in ([0, \infty)$ denotes the size of the epoch-specific reward pool.\footnote{Reward pool is a distinct pool holding accrued rewards generated in PoEL protocol for securing the network, and optimally distributing it to LPs, based on pre-defined rules.}\\
    \\
    The rules-based mechanism driving the reward distribution seeks to optimise allocation by granting preferential rewards to relatively more desirable assets, thus maintaining capital efficiency in the protocol. This reward distribution policy is further informed by the collateralisation rate, which seeks to help incorporate risk management dynamics, into the reward distribution policy.
    
    \begin{itemize}
        \item \textbf{Budget Distribution}: PoEL protocol's budget distribution mechanism seeks to allocate the available budget across multiple epochs ($R_e^i \leq RP_e$), where $R_e^i$ represents the optimal epoch-specific reward for LPs of i-th CDA and $RP_e$ is the maximum budget available for distribution in that epoch, with the intent of striking a balance between capital efficiency and boosting incentive allocation in epochs with elevated user demand.
        
        \item \textbf{Dynamic Interest Rate}: PoEL quantifies a fee on borrowers of the native staking token, which depends on the quality of the asset used for collateral, and serves the purpose of efficiently distributing reward budget, higher quality assets which have a higher demand as operating capital in a CDA are given preferential rates. This rate, represented by $I^{\tilde{X}^i}_e \in \mathop{\mathbb{R}}$, is specific to an LP token and epoch, and computed as a percentage of the distributable rewards for contributors of that LP token as collateral($R^{\tilde{X}^i}_e$), that is calculated as follows:

        \begin{equation}
            R^{\tilde{X}^i}_e =  R^i_e \cdot \frac{L^{\tilde{X}^i}_e} {L_e}.
        \end{equation}
        
        \noindent
        Similarly, the nominal interest payable ($IP^{\tilde{X}^i}_e$) is calculated as follows:
        \begin{equation}
            IP^{\tilde{X}^i}_e = R^{\tilde{X}^i}_e \cdot I^{\tilde{X}^i}_e
        \end{equation}
  
        \item \textbf{Collateralisation Rate}: Instead of using a static amount of collateral required to borrow each unit of the native staking token, the protocol sets this rate dynamically, as another tool to assert its pre-specified principles\footnote{Principle 2 requires that $\rho^{X^i_{LP}}_e \geq 1$.} and objectives. Furthermore, the PoEL protocol aims to accommodate any fluctuations in the combined value of assets used as collateral, by modifying the loaned amount as follows:
        
        \begin{equation}
            \Delta L^{\tilde{X}}_e = \frac{S^{\tilde{X}^i}_e}{\rho^{\tilde{X}^i}_e} \cdot P^{\tilde{X}^i}_e - L^{\tilde{X}}_{e-1}.
        \end{equation}
    
        \noindent
        Note that in the event of an adverse price change which leads to curtailment of the loaned amount ($\Delta L^{\tilde{X}}_t < 0$), ownership of the loaned asset is transferred to the PoEL protocol. It can then be reassigned to another LP capable of providing collateral, or it may trigger the unstaking process to commence the withdrawal of tokens from the PoS scheme and return them to the network \footnote{Note that if $\Delta L^{\tilde{X}^i}_e$ is a positive value this does not necessarily assume that the system will extend new loans in that epoch, based on the staking dynamics of the implementing blockchain the extension of new loans in PoEL protocol might be subject to frequency constraints.}. Here, $\rho^{\tilde{X}^i}_e$ represents the collateralisation rate of a specific asset.
   \end{itemize}

   \noindent
    Essentially, this reward distribution mechanism aims to harmonise individual risk management policies with their respective objectives. It is important to note that the protocol mitigates credit risk by controlling not only the collateral but also the loaned tokens allocated because the borrowed tokens are never accessible to the borrowers.\\
    \\
    Finally, rewards are optimally allocated to the different LP token reserves and can be withdrawn by the LPs with the LP tokens. 

\subsection{Operational Integrity}
    Operationally, LPs deposit LP tokens in LP token reserves, which are subsequently delegated to validators in the form of a loan, thereby promoting PoEL's objective to use liquidity submitted in liquidity pools to secure the network. However, this has to be achieved in a manner where the interests of the protocol are aligned with those of its agents.\\
    \\
    This alignment is achieved as follows: 
    
    \begin{itemize}
        \item \textbf{Validator Due Diligence}: The protocol incentivises LPs to conduct their due diligence before selecting the validator as a prospective candidate for the delegation of their assets. One method of achieving this is by slashing the collateral of LPs who delegate to misbehaving validators. If penalised, on a ceteris paribus basis, the change of the total lent amount of the native staking token delegated to a particular validator $v$ ($\Delta L_{e_v}$) can be calculated as follows: 
        
        \begin{equation}
            \Delta L_{e_v} = L_{{e-1}_v} \cdot (1-\varpi_e),
        \end{equation}
       
        \noindent
        where $\varpi_e$ is the slashing rate regulated by the adopting network's PoS mechanism, or set independently as a parameter in PoEL; and $L_{{e-1}_v}$ represents the loaned amount of native staking tokens delegated to v-th validator in the previous epoch. On a ceteris paribus basis, once slashed, an LP token reserve balance  ($S^{\tilde{X}^i}_{e_v}$) would change as follows:
        \begin{equation}
            S^{{\tilde{X}^i}}_{e_v} = S^{{\tilde{X}^i}}_{{e-1}_v} -  \frac{\frac{S^{\tilde{X}^i}_{{e-1}_v} \cdot P^{\tilde{X}^i}_e}{\sum^g_{j=1}S^{j}_{{e-1}_v} \cdot P^{j}_e} \cdot \Delta L_{e_v}} { {P^{\tilde{X}^i}_e}},
        \end{equation}

        \noindent
        where $g \in \mathop{\mathbb{I}}^+$ is the highest index of LP tokens submitted to the LP token reserve that are mapped to validator $v$'s address (the subject of slashing); and $P^{{\tilde{X}}^i}_{e}$ is the price of the LP token $\tilde{X}^i$ in e-th epoch.\\
        \\
        Slashing thus serves as a deterrent against LPs who might strategically delegate to malicious validators, possibly intending to compromise network quality, or among other effects, cause a depreciation in the price of the network's native staking token.

        \item \textbf{Liveness}: The network might stall if a sufficient number of validators breach the core principle of minimum staking requirement, leading to their ejection, which if happening concurrently (or cascadingly) could pose a risk to the network's liveness (Risk 3). This issue must be systematically addressed while considering the unique design and economic characteristics of each adopting blockchain network. We seek to accomplish this by quantifying a ceiling of the amount of loaned native staking tokens that are available to be delegated to validators, whilst considering: 
        \begin{itemize}
            \item Minimum viable number of nodes required for the adopting blockchain network to remain operational.
            \item Minimum staking requirement enforced in the adopting blockchain network.
        \end{itemize}

        \noindent
        Additionally, a range of other network-specific factors must be considered. These are generally incorporated into a function used to quantify the loanable ceiling ($C_{e_v}$) for each validator, ensuring that $C_{e_v} \geq L_{e_v}$.
    
        \item \textbf{Diversification}: PoEL incentivises diversification of assets comprising the collateral base, to manage the resulting impact of volatility stemming from assets locked in the network. In essence, if the asset base had n-many different assets\footnote{Under the presumption that the underlying multivariate distribution characterising the basket of assets maintains the relevance and applicability of the covariance matrix.}, then the following can be written for the portfolio-level variance.

        \begin{equation}
            \sigma^2_p = \sum_{b=1}^{n} \sum_{j=1}^{n} w_b \cdot w_j \cdot \rho_{b,j} \cdot \sigma_b \cdot \sigma_j,
        \end{equation}
        
        \noindent
        where $\sigma_b$ and $\sigma_j$ are individual asset volatility, $\rho_{b,j}$ is the correlation coefficient between two assets, $\sigma^2_p$ is the portfolio variance, $w_b$ and $w_j$ are asset-specific weights.

        \item \textbf{Tenure}: PoEL incentivises LPs to lock their assets in a CDA for a longer period, to effectively help smooth out the value of the locked assets, thereby, promoting the long-term participation of LPs and stabilising the system.
    \end{itemize}

    \noindent
    In the forthcoming sections, we elaborate on some of the some of the levers used to assert the network's operational integrity.
    
\subsubsection{Dynamic Collaterisation Rate}
    The protocol quantifies collateral, which is the amount of an asset that must be staked by LPs to gain access to borrowed native staking tokens to perform work themselves, or delegate it to another validator. To ensure that despite evolving dynamics, the protocol's state is either at, or always progressing towards meeting its risk-conscious economic objectives, the collateralisation rate is quantified dynamically for individual assets. This rate is dynamically adjusted in keeping with the following objectives, which are continuously tested individually, with resulting changes that are required applied cumulatively.
    
    \begin{itemize}
        \item \textbf{Qualification}: A decentralised governance mechanism may be used to enforce a list of criteria, e.g. minimum liquidity, maximum volatility, decentralisation, etc., that must be satisfied before an asset is admissible to the universe acceptable as collateral.
    
        \item \textbf{Risk Homogenisation}: We standardise the varying quality of assets provided by LPs as collateral, enabling the protocol to homogenise the risk-adjusted desirability of an asset, leading to the following condition to be held:
        \begin{equation}
            \rho^{\tilde{X}^i} \propto (\xi^{\tilde{X}^i} - \bar{\xi})^y \forall y>0
        \end{equation}

        \noindent
        where $\xi$ represents an abstraction of all measures of asset-specific risk; and $\bar{\xi}$ is the representative risk of all assets in the basket, say the median or mean.

        \item \textbf{Diversification}: The quality of collateral is crucial, yet the pursuit of specific assets must be weighed against diversification to mitigate both known and unknown risks. We calculate the collateralisation rate for an asset ($\rho^{{\tilde{X}}^i}_e$) as follows:

       \begin{equation}
          \rho^ {\tilde{X}^i}_e = \rho_{min} +  \Delta \rho^{\tilde{X}^i}_e,
        \end{equation}
        
        \noindent
        where $\rho_{min}$ is the minimum collateralisation rate, and $\Delta \rho^{{\tilde{X}}^i}_e$ can be calculated as:

        \begin{align}
            \Delta \rho_e ^ {\tilde{X}^i} &= \eta^{\tilde{X}^i}_e \cdot \left( e^{\left(\chi + b \cdot \text{sign}\left(\frac{w^{\tilde{X}^i}_e - w^{\tilde{X}^{i\ast}}_e}{w^{\tilde{X}^{i\ast}}_e}\right) \right) \cdot \left| \frac{w^{\tilde{X}^i}_e - w^{\tilde{X}^{i\ast}}_e}{w^{\tilde{X}^{i\ast}}_e} \right| } - 1 \right), & \chi \geq b.
        \end{align}

        \noindent
        Here, $\eta^{\tilde{X}^i}_e$ is the asset $\tilde{X}^i$ specific risk scaling factor in the e-th epoch, which seeks to control the tradeoff between the deviation of an asset's weight in the asset-basket (risk) from the target rate\footnote{For simplicity, the risk scaling factor would be initially set to be the same for all assets, i.e. $\eta^{\tilde{X}^i}_e = \eta_e \forall \tilde{X},i$.}; $\chi$ adjusts sensitivity to weight deviations from targets by determining the steepness of the exponential curve, allows for control over how expediently the collateralisation rate responds to changes in asset weights; $b$ adjusts the steepness of the curve dependent on whether there is positive or negative deviation, intended to compel expedient action of asset providers; $w^{\tilde{X}^{i^{\ast}}}_e$ represents target weights of asset $\tilde{X}^i \neq 0$\footnote{$w^{\tilde{X}^{i^{\ast}}}_t = 0$ would mean that an asset cannot be admitted as collateral.}; and ${w^{{\tilde{X}}^i}_e}$ represents the weight of the asset in the basket in e-th epoch. This approach means that an excess or deficit of a specific asset influences the collateralisation rate. Specifically note that for all practical purposes $\chi = b$, assuming that security considerations mean that system parameters cannot be changed as frequently.\\
        \\
        Explicitly, it should be observed that the exponential function within the expression encapsulates a non-linear reaction to deviations in weight, which is suitably contingent upon both the magnitude and direction of the divergence of weights from the predefined target.\\
        \\
        Now, the target asset-specific mix $w^{\tilde{X}^{i^{\ast}}}_e \in \mathop{\mathbb{R}}^{+}$ is set off-chain using the following optimisation problem: 
        \begin{equation*}
            \underset{\mathbf{w^\ast}}{\text{Minimise}} \sum_{b} \sum_{j} w_b w_j \sigma_b \sigma_j \rho_{bj},
        \end{equation*}
        subject to,
        \begin{equation*}
            w_{NST} \geq w_{NST}^{\ast}
        \end{equation*}
        \begin{equation*}
            \sigma_{b \neq NST} \leq \lceil \sigma \rceil,
        \end{equation*}
        \begin{equation*}
            \frac{1}{1 - CI} \int_{-\infty}^{VaR} r_p\cdot dF(r_p) \leq ES^\ast,
        \end{equation*}
        
        \noindent
        where $\mathbf{w^\ast}$ represents the target weights of different assets in the portfolio; $\sigma_i \in \mathop{\mathbb{R}}^{+}$ and $ES^\ast \in \mathop{\mathbb{R}}^{+}$ is variance of an individual asset with weight $w$ and expected shortfall limits of the overall asset mix of the portfolio\footnote{This model posits that by imposing a constraint on the expected shortfall of the asset portfolio, the PoEL protocol can effectively minimise financial risks associated with systemic vulnerabilities, such as sudden and significant asset price declines.}, respectively; $\lceil \sigma \rceil$ represents the pre-defined ceiling on the volatility of collateral portfolio; $F(R)$ is the cumulative distribution of returns ($r_p = \sum_{j=1}^{n} w_j r_j$) of assets comprising the portfolio; $CI$ is the confidence level; $VaR$ represents the value at risk; and $w^{\ast}_{NST} \in \mathop{\mathbb{R}}^{+}$ is the target weight allocated to the native staking token in the asset basket, based on the capital staked by capital providers directly in the native staking token, excluding the native staking token that has been borrowed by LPs.\\
        \\
        This problem is comprised of quadratic objectives (portfolio variance) and linear and non-linear constraints (weight and risk constraints). Whilst a standard Quadratic Programming solver could have efficiently tackled a convex optimisation problem comprised of a quadratic objective function and linear constraints, the presence of non-linear constraint in the expected shortfall and inherent uncertainty in variables involved in the optimisation statement, a Stochastic Programming solver would be pertinent, as naturally if there wasn't any inherent uncertainty, a Conic Optimisation solver could have also sufficed. We only briefly state these alternatives, as the precise methodology to be adopted for solving this off-chain is beyond the scope of this work.
        
       \item \textbf{Unstaking Time}: The unstaking time, as defined previously, is implemented as a security measure in PoS blockchain networks, but can expose the PoEL protocol to Risk 2, i.e. if the length of an epoch is shorter than the unstaking time, an adverse shift in asset price may lead to a scenario where $L_e > \sum^g_{j=1} S^j_e \cdot P^j_e$, which is even applicable if the ownership of the loaned NSTs transfers from the user to the protocol. Since this risk cannot be directly mitigated, we seek to manage the impact if this risk were to emerge, through an upper bound on the variance of the portfolio of assets that are accepted as collateral in the system.

        \item \textbf{Operational Robustness}: To avoid scenarios where an agent attempts to borrow a significant enough portion of the native staking token to attempt a stake centralisation attack, we seek to enforce a maximum threshold loaned ownership of native staking tokens by LPs. This is achieved by asserting that the weight of native staking tokens directly staked in the PoS (not borrowed from the network) is not below a particular threshold on the loaned asset, i.e. 
        \begin{equation*}
            w_{NST} = 1-\frac{L_e}{\text{Total staked capital in the native staking token}_e} \geq w^\ast_{NST}
        \end{equation*}
        
        \noindent    
        This ensures that the PoEL protocol maximally mitigates Risk 5, and this threshold weight could be used to quantify the amount of native staking tokens which can be borrowed in any epoch($\mathcal{T}_e$), as follows: 
        \begin{equation*}
            \mathcal{T}_e = \text{Total staked capital in the native staking token}_e \cdot ( 1 - w^\ast_{NST})
        \end{equation*}
    \end{itemize}

    \noindent
    In essence, the dynamic collateralisation rate requiring a higher (or lower) collateral amount on a relative basis to another asset, imposes an opportunity cost on the LP seeking to borrow native staking tokens, thereby, enabling the PoEL protocol to meet its stated risk diversification objective.\\
    \\
    To ensure that the system fulfils the operational robustness objective described above, the collaterisation rate needs to dynamically change to accommodate the threshold $\mathcal{T}_e$. As such, we now consider the dynamics of the system when the collateral is comprised of three assets, say, $\tilde{X}^i$, $\tilde{Z}^i$ and $\tilde{Y}^i$, such that the amount of borrowed tokens can be calculated as follows:

    \begin{equation}
        L_e = \sum_{h \in \{\tilde{X}^i, \tilde{Y}^i, \tilde{Z}^i\}} \frac{S^h_e \cdot P^h_e}{\rho^h_e}.
    \end{equation}

    \noindent
    If the user continues to add more assets to the LP token reserves, or if the price of an asset increases, the amount of the loaned native staking token could exceed the maximum threshold $\mathcal{T}_e$. For such cases, we introduce the variable $m$, which is applied as a multiplier to the collateralisation rate, aiming to increase the collateralisation rate of all assets to a point that would ensure the total borrowed amount $L_e$ does not exceed the predefined threshold $T_e$.

    \begin{equation}
        \sum_{h \in \{\tilde{X}^i, \tilde{Y}^i, \tilde{Z}^i\}} \frac{S^h_e \cdot P^h_e}{\rho^h_e \cdot m} = \mathcal{T}.
    \end{equation}

    \noindent
    We can now rearrange the preceding equation to calculate $m$, as follows:
     \begin{equation}
         m = \frac{\sum_{h, j, k \in \{\tilde{X}^i, \tilde{Y}^i, \tilde{Z}^i\}, h \neq j \neq k} S^h_e \cdot \rho^j_e \cdot \rho^k_e \cdot P^h_e}{\prod_{h \in \{\tilde{X}^i, \tilde{Y}^i, \tilde{Z}^i\}} \rho^h_e \cdot \mathcal{T}}
     \end{equation}

\subsubsection{Locking Time}
    The PoEL protocol allows LPs to choose the duration for which they wish to lock their assets in LP token reserves, a period henceforth referred to as the \textit{locking time}. Concurrently, it provides incentives to encourage sticky capital, thereby promoting the network's economic security and stability. This is accomplished through the specified incentive for Target 1, where the final reward received by an LP/borrower $l$ in epoch $e$ is determined by a function, which we reiterate below:
    
    \begin{equation}
        R_e^l = R_{e}^{l^{\min}} + \frac{R_e^{l^{\max}} - R_e^{l^{\min}}}{(1 + e^{-k (e^l - e_{\text{mid}}^l)})^\nu}
    \end{equation}

    \noindent
    where $e^l$ represents the number of epochs for which l-th LP's capital is locked, $R_e^l$ represents the associated incentive, $R_e^{l^{\max}}$ represents the maximum incentive available to distribute to a particular LP at a particular epoch, $R_e^{l^{\min}}$ is the minimum incentive (a portion of $R^{l^{\max}}_e$), $k$ is a scaling factor that adjusts the steepness of the sigmoid curve, $e_{\text{mid}}$ is a mid-point value around which the transition from the minimum to maximum incentive starts to become significant, and $\nu$ is a new parameter that adjusts the curve's shape, providing additional control over how rapidly the incentive increases as the locking time approaches and surpasses mid-point.\\
    \\
    The maximum reward itself ($R^{l^{\max}}_e$) is computed as follows: 
    \begin{equation}
        R^{l^{\max}}_e = \text{StakingRewardsDistributed} - \text{InterestCharged}
    \end{equation}

    \noindent
    Having discussed operational aspects, we now discuss how staking rewards and interest rates are used to achieve the protocol's objectives. 

\subsection{Financial Management}
    To assert the adopting network's objectives, encourage a risk-aware choice for LPs, and optimally use the budgeted incentives allocated to attract liquidity, the protocol complements income generated from CDA-specific fees with the allocation of preferential rewards (generated from securing the network) and variable lending fees. This approach results in a combination of risks and rewards (preferential incentives) which aid the PoEL protocol in achieving its stated objectives.

\subsubsection{Dynamic Reward Allocation}\label{sec:dra}
    Rewards are accumulated for LPs who delegate their capital to validators, thus enabling them to meet staking requirements and assert the economic security of the network. These rewards are kept in a combined reward pool, as shown in Fig.\ref{fig:helicopterview}. The pool's resources, reflecting its health, finance the preferential incentive budget. This budget is carefully distributed to support the protocol's objectives, notably Target 3, which focuses on enhancing capital efficiency and fostering user adoption. The allocation of this budget for preferential incentives in each epoch is determined by the following optimisation problem, which we aim to solve using smart contracts by tracing user on-chain behaviour:
    \begin{equation}
    \begin{aligned}
        & \underset{\mathbf{R}^i_{1:\epsilon}}{Maximise} \mathop{\mathbb{E}}[E^i_{1:\epsilon}(\mathbf{R}^i_{1:\epsilon})]\\
        & \textbf{s.t.}\\
        & R^i_{{min}_e} \leq R^i_{e} \leq RP_e \: \text{and} \:  \sum^{\epsilon}_{o=1}  R^i_{o} = \sum^{\epsilon}_{o=1} SR_o
    \end{aligned}
    \end{equation}   

    \noindent
    where $E^i_{1:\epsilon}$ is the capital efficiency of i-th CDA \footnote{The described mechanism assumes that only a single CDA is involved in the PoEL program in any epoch, and the system does aim to optimise the reward distribution between different CDAs.}, which is defined below; $RP_e$ is the size of the reward pool \footnote{The size of the reward pool could be calculated based on total earned staking rewards from PoS for the borrowed native staking token and distributed rewards to LPs $RP_e = \sum^{e}_{o=1} SR_o - \sum^{e}_{o=1}  R^i_{o-1}$}; $R^i_{{min}_e}$ is the minimum staking reward that needs to be distributed in the epoch $e$ as part of the liquidity bootstrapping program and is based on a system parameter; $\epsilon$ is the maximum number of epochs for which the liquidity mining program is operational\footnote{The period for which the incentive programme is operational may be longer than the period during which staking rewards are distributed to the reward pool, i.e., $\epsilon \geq \Omega$, where $\Omega$ is the number of epochs during which there is a distribution of the staking rewards to the reward pool. In other words, this is intended to reiterate that the duration of the incentive programme is distinct from the duration for which borrowed NSTs could be staked in the network, i.e., the system will continue distributing rewards from the pool to LPs as long as $RP_e \geq 0$, even after the end of the staking reward distribution period}; $\mathbf{R}^i_{1:\epsilon} = \{R^i_1, R^i_2, \dots, R^i_{\epsilon}\}$ is a vector of total reward budget distributed to LPs in a particular CDA across epochs; and $SR_e \in [0, \infty)]$ is the total staking reward earned and distributed to reward pool for borrowed native staking tokens.\\
    \\
    Now, to enable us to track the efficiency of any rewards distributed to attract liquidity, we introduce a new measure - Capital Efficiency Rate.

    \begin{definition}[Capital Efficiency Rate]
        The capital efficiency rate for a specific CDA measures the proportion of capital submitted by LPs as operating capital to liquidity pools of a CDA, which has been utilised by the user, and quantified as follows:
        \begin{equation}
           E^i_e = max(\frac{\mathcal{L}^i_{LP_e} - \mathcal{L}^i_e}{\mathcal{L}^i_{LP_e}},0)
        \end{equation}
    
        \noindent
        where $\mathcal{L}_e^i \in [0, \infty)$ is the (nominal) capital available in the participating\footnote{For simplicity, we assume that all liquidity pools of a CDA are participating in the PoEL incentivisation program.} liquidity pools of i-th CDA in e-th epoch; and $\mathcal{L}^i_{LP_e} \in \mathop{\mathbb{R}}^{+}$ represents the (nominal) LP capital in e-th epoch, which has been deposited in the pools.
    \end{definition}

    \noindent
    An alternative definition of the preceding measure can be calculated to measure the productivity of distributed rewards for the fees that those rewards generate, i.e. $E^i_e = \frac{F^i_e}{R^i_e}$, where the $F^i_e$ is the fees earned in nominal terms during the epoch by the CDA(as a result of its usage), and $R^i_e$ is the rewards distributed with the epoch to LPs of CDA.\\
    \\
    Now this measure can be generalised to different DeFi applications, which enables us to measure if a CDA has the potential to be sustainable beyond the period when additional incentives are provided to attract liquidity, which is largely aimed at breaking the interdependent causal loop referenced in the introductory sections. This potential can be measured by its ability to generate a high enough capital efficiency, which will help maximise the financial value of the transactional volume.\\
    \\
    Now, since there is a finite amount of budget available (staking rewards accumulated in the reward pool) to distribute as incentives, we front-load incentives to help kick-start the system in its nascent stages, and when it is experiencing high capital efficiency rates - to meet user demand. Therefore, the reward ($R^i_e$) is quantified as follows: 
    \begin{equation}\label{eq:rewards}
        R^i_e = min(SRR_e \cdot L_e^i \cdot (max(\zeta,1 - \Theta \cdot (\mathcal{E}^{i,m^{\ast}}_e - \mathcal{E}_{e}^{i,m})^c)), RP_e),
    \end{equation}

    \noindent
    where $m$ represents the number of epochs over which the moving average of the capital efficiency ($\mathcal{E}^{i,m}$) is calculated; $RP_e$ is the size of epoch-specific reward pool; $\mathcal{E}^{{i,m}^{\ast}}_e$ is the target capital efficiency rate in a window $m$; $SRR_e$ is the staking reward rate of the adopting blockchain in $e$-th epoch; $\Theta$ and $c \forall c\%2 \neq 0$ are parameters governing the relationship between the deviation from target capital efficiency rate and the reward distributed; $\zeta$ is a parameter which defines the smallest portion of staking rewards that need to be distributed to bootstrap the liquidity of CDA.\\
    \\
    We propose an algorithm (Appendix \ref{apx:targetcer}) to optimise the protocol's target (moving average) capital efficiency rate ($\mathcal{E}^{{i,m}^{\ast}}_e$), to enhance a CDA's average capital efficiency. The algorithm tracks changes in the moving average capital efficiency rate over two distinct time windows, $m \in [e_1, e_2]$ and $n \in [e_3, e_4]$, where $e_1 > e_3$, $e_2 = e_4$, and $e_4 - e_3 > e_2 - e_1$. The algorithm initiates by setting an initial target capital efficiency rate ($\mathcal{E}_0^{{i,m}^{\ast}}$) and defining lengths for the short ($m$) and long ($n$) time windows. Global parameters, namely $\Upsilon$ regulate the speed of negative adjustments, and $\psi$ regulates the speed of positive adjustments. It calculates the first ($\frac{\partial \mathcal{E}^{i,m}}{\partial e}$) and second ($\frac{\partial^2 \mathcal{E}^{i,m}}{\partial e^2}$) derivatives of the moving average of capital efficiency rate over last epoch. The decision logic involves increasing or decreasing the target rate ($\mathcal{E}^{{i,m}^{\ast}}_e$) based on the direction and acceleration of these derivatives. If both windows' first derivative indicate an efficiency increase, the target rate decreases; conversely, it increases if there's a decrease in efficiency. The extent of adjustment is determined by the alignment of the second derivatives in both windows, with full adjustments made when both windows corroborate the acceleration or deceleration of the trend, and reduced adjustments otherwise. The algorithm outputs the adjusted target capital efficiency rate, effectively balancing short-term responsiveness with long-term trend analysis.\\
    \\
    The dynamic formulation presented in Algorithm 1 enables the protocol to use a direct measure of the rate of change, as opposed to its lagged and vulnerable (e.g. to whipsawing markets) proxies like the crossover of moving averages. The use of both long and short windows, first and second derivatives, enables the protocol to be more sensitive to short-term trends, but to accomplish it more effectively in the context of historical data. However, note that because we operate in a discrete-time world, where the capital efficiency rate is measured in equally spaced intervals of one epoch we would be required to approximate the first and second derivatives, as follows, for example:
    
    \begin{equation*}
        \frac{\partial \mathcal{E}^{i,m}}{\partial e} \approx {\mathcal{E}^{i,m}_{e} - \mathcal{E}_{e - 1}^{i,m}},
    \end{equation*}
    
    \begin{equation*}
        \frac{\partial^2 \mathcal{E}^{i,m}}{\partial e^2}  \approx \mathcal{E}_{e}^{i,m} + \mathcal{E}_{e - 2}^{i,m} - 2\mathcal{E}_{e-1}^{i,m}.
    \end{equation*}
    
\subsubsection{Variable Lending Fee}
    Complementing the dynamic reward allocation scheme presented in the preceding section, the PoEL protocol uses interest rates to enable it to meet its stated objectives. This interest ($I^{\tilde{X}^i}_e$) is calculated as a portion of total rewards distributable to LPs, which is dynamically calculated to optimise the distribution of rewards between different LPs who submitted different LP tokens and thereby to maximise aggregate capital efficiency. This rate is modulated through a smart contract or a combination of on-chain/off-chain trustless computation logic, aiming to solve a single-objective optimisation problem. It is constrained by a reward budget $R_e$ for the $e$-th epoch, seeking to maximise the total capital efficiency of the $i$-th CDA in the $e$-th epoch. This efficiency has a monotonic relationship with the fees earned by a CDA.
    
    \begin{equation}
    \begin{aligned}
        & \underset{\mathbf{I}_e}{maximise} E^i_e(\mathbf{I}_e)  \\
        & \textbf{s.t.}\\
        & \sum^g_{j=1} R^{j}_{e} \leq R^i_e
    \end{aligned}
    \end{equation}

    \noindent
    where $g$ is the highest index of LP tokens from different pools of i-th CDA that are submitted to the PoEL protocol; and $\mathbf{I}_e$ is a vector of interest rates charged from rewards distributable for staked NST that has been borrowed using different LP tokens as collateral; $R^j_e$ represents the staking rewards distributable to LP token reserves for a staked native staking token that has been borrowed using $j$-th asset as collateral. Logically, solving the above optimisation problem necessitates the system to incentivise pools experiencing increased demand, while reducing incentives for less popular pools. Therefore, we now quantify the capital efficiency rate ($E^{i^X}_e$) for an asset pool within a CDA:
    
    \begin{equation}
        E^{i^X}_e = max(\frac{\mathcal{L}^{i^X}_{{LP}_e} - \mathcal{L}^{i^X}_e}{\mathcal{L}^{i^X}_{LP_e}},0),
    \end{equation}

    \noindent
    where $\mathcal{L}^{i^X}_e$ is the (capital) liquidity available in asset pool $X$ of the CDA $i$ at the epoch $e$; and $\mathcal{L}^{i^X}_{{LP}_e}$ is the liquidity of the asset $X$ in CDA $i$ that has been submitted by LPs \footnote{Similar to the alternative definition of the capital efficiency rate provided for CDAs, an alternative definition of the same measure for a particular asset pool is: $E^{i^X}_e = \frac{F^{i^X}_e}{R^{\tilde{X}^i}_e}$, where $F^{i^X}_e$ is the fees earned from utilisation of working capital of asset pool $X$ of i-th CDA during the epoch $e$.}. It can be observed that the aforementioned formulation essentially provides a point-in-time snapshot of the capital efficiency rate, which may leave room for malicious users to manipulate key measures, therefore, instead of using a point-in-time measure, we seek to calculate the capital efficiency rate over a period of time (look-back window of length - $m$ epochs), and denote it with $\mathcal{E}^{{i,m}^X}_{e}$.\\
    \\
    To quantify the optimal interest rate $I^{\tilde{X}^i}_e \in \mathbb{R}$ for various assets, we identify the asset pool within a CDA which exhibits the highest and lowest (moving average) capital efficiency, designated as $\mathcal{E}_e^{max}$ and $\mathcal{E}_e^{min}$, respectively. On a ceteris paribus basis, allocated rewards are positively correlated to the capital efficiency in the system, which leads to a scenario where the asset pool with the lowest capital efficiency ($\mathcal{E}_e^{min}$) attracts the lowest reward, and the asset with the highest capital efficiency ($\mathcal{E}_e^{max}$) attracts the highest reward. We use these dynamics to scale allocated rewards, using the weight $\mathcal{W}^{\tilde{X}^i}_e$, with increasing capital efficiency, as follows:

   \begin{equation}
        \mathcal{W}^{\tilde{X}^i}_e = \lfloor{\mathcal{W}} \rfloor + (\lceil{\mathcal{W}}\rceil-\lfloor{\mathcal{W}}\rfloor)  \cdot ( 1- (\frac{\mathcal{E}_e^{max} - \mathcal{E}_e^{{i,m}^X}}{\mathcal{E}_e^{max}- \mathcal{E}_e^{min}})^\mathcal{K})
    \end{equation}

    \noindent
    where $\lfloor{\mathcal{W}} \rfloor$ and $\lceil{\mathcal{W}} \rceil$ are minimum and maximum weights attached to assets; $\mathcal{K}$ is a system parameter governing the convexity of relationship between weights and the capital efficiency rate; and $\mathcal{G}$ is the factor which governs the desirability of a higher capital efficiency asset in the portfolio, such that $\lceil{\mathcal{W}} \rceil = \lfloor{\mathcal{W}} \rfloor \cdot (1 + \mathcal{G})$.\\
    \\
    Now, the rewards distributable ($DR^{{\tilde{X}}^i}_e$) to an LP's token reserves for an asset ($\tilde{X}^i$) in an epoch is calculated as follows:
    
    \begin{equation}
        DR^{{\tilde{X}}^i}_e= \frac{L^{{\tilde{X}}^i}_e \cdot \mathcal{W}^{\tilde{X}^i}_e}{\sum^g_{j=1} L^j_e \cdot \mathcal{W}^j_e} \cdot R^i_e,
    \end{equation}

     \noindent
     where $\sum_{\tilde{X}^i} DR^{{\tilde{X}}^i}_e$ represents the distributable rewards across all assets.\\
     \\
     Now, we can also calculate the applicable interest rate charged for each asset ($I^{\tilde{X}^i}_e \in \mathbb{R}$), as follows:

    \begin{equation}
        I^{\tilde{X}^i}_e =  1-\frac{DR^{{\tilde{X}}^i}_e}{R^{\tilde{X}^i}_e}.
    \end{equation}

\section{Conclusion}\label{sec:conclusion}
    In this work, we have presented the concept of Proof of Efficient Liquidity (PoEL) protocol, which is designed for PoS blockchain infrastructures with intrinsic DeFi applications, seeking to support sustainable liquidity bootstrapping and network security. We have reviewed existing literature (Sec. \ref{sec:litreview}), in the context of which the protocol's governing principles (Sec. \ref{sec:principles}), and the proposed protocol architecture (Sec. \ref{sec:protocol}) can be understood.\\
    \\
    In forthcoming revisions, we will formally prove that the proposed protocol architecture meets its stated objectives, provide case studies through agent base simulation, and pave the path for expanding the protocol to an algorithmic model for interest rate calculations using LPs on-chain reputation scores and loss modelling, enabling us to price and manage varied risks stemming from liquidity bootstrapping. Finally, we will also explore decentralised automation services to optimise capital allocation and harmonise incentive schemes, in the context of competing incentive programs running on the same blockchain network, albeit for different CDAs.

\appendix
\section*{Appendices}
\section{Algorithm 1: Target Capital Efficiency Rate}\label{apx:targetcer}
    \begin{algorithm}[H]
    \caption{Target Capital Efficiency Rate}
    \begin{algorithmic}[1]
        \Require Derivative accumulations $D^{(1)}_m, D^{(1)}_n, D^{(2)}_m, D^{(2)}_n$
        \State $\mathcal{E}_0^{{i,m}^{\ast}} \leftarrow$ {Initial target capital efficiency rate}
        \State $m, n \leftarrow$ {Lengths of the short and long time windows, respectively}
        \State $\Upsilon, \psi \leftarrow$ {Global parameters regulating the speed of adjustments}
        \State $q^{'}, q^{''} \leftarrow$ {Parameters governing the extent to which disagreement in second derivatives over long/short horizon augments the scale of change}
        % \State $Y \leftarrow$ {Global parameter indicating market response time.}
        \State $\mathcal{B}_{lower}^{(1)} \leftarrow$ {Lower bound for the first derivative}
        \State \textbf{Calculate} $\frac{\partial \mathcal{E}^{i,m}}{\partial e}, \frac{\partial^2 \mathcal{E}^{i,m}}{\partial e^2}, \frac{\partial \mathcal{E}^{i,n}}{\partial e}, \frac{\partial^2 \mathcal{E}^{i,n}}{\partial e^2}$
        \If{$\mathcal{B}_{lower}^{(1)} \leq \max(\left|\frac{\partial \mathcal{E}^{i,m}}{\partial e}\right|, \left|D^{(1)}_m\right|)$ \textbf{and}}
            \State $\mathcal{B}_{lower}^{(1)} \leq \max(\left|\frac{\partial \mathcal{E}^{i,n}}{\partial e}\right|, \left|D^{(1)}_n\right|)$ \textbf{and}
            \If{$\max(\frac{\partial \mathcal{E}^{i,m}}{\partial e}, D^{(1)}_m) > 0$ \textbf{and} $\max(\frac{\partial \mathcal{E}_n}{\partial e}, D^{(1)}_n) > 0$}
                \If{$\max(\frac{\partial^2 \mathcal{E}^{i,m}}{\partial t^2}, D^{(2)}_m) \geq 0$ \textbf{and} $\max(\frac{\partial^2 \mathcal{E}^{i,n}}{\partial e^2}, D^{(2)}_n) \geq 0$}
                    \State $\mathcal{E}^{{i,m}^\ast}_{e}  = \mathcal{E}^{{i,m}^\ast}_{e} - \Upsilon$
                \Else
                    \State $\mathcal{E}^{{i,m}^\ast}_{e}  = \mathcal{E}^{{i,m}^\ast}_{e}  - \frac{\Upsilon}{\max({q^{'} \cdot\left|\max\left(\frac{\partial^2 \mathcal{E}^{i,m}}{\partial e^2}, D^{(2)}_m\right) - \max\left(\frac{\partial^2 \mathcal{E}^{i,n}}{\partial e^2}, D^{(2)}_n\right)\right|}, 1)}$
                \EndIf
            \ElsIf{$\max(\frac{\partial \mathcal{E}^{i,m}}{\partial e}, D^{(1)}_m) < 0$ \textbf{and} $\max(\frac{\partial \mathcal{E}^{i,n}}{\partial e}, D^{(1)}_n) < 0$}
                \If{$\max(\frac{\partial^2 \mathcal{E}^{i,m}}{\partial e^2}, D^{(2)}_m) \leq 0$ \textbf{and} $\max(\frac{\partial e^2 \mathcal{E}^{i,n}}{\partial e^2}, D^{(2)}_n) \leq 0$}
                    \State $\mathcal{E}^{{i,m}^{\ast}}_{e} = \mathcal{E}^{{i,m}^\ast}_{e}  + \psi$
                \Else
                    \State $\mathcal{E}^\ast_{j_{t}} = \mathcal{E}^{{i,m}^\ast}_{e}  + \frac{\psi}{\max({q^{''} \cdot\left|\max\left(\frac{\partial^2 \mathcal{E}^{i,m}}{\partial e^2}, D^{(2)}_m\right) - \max\left(\frac{\partial^2 \mathcal{E}^{i,n}}{\partial e^2}, D^{(2)}_n\right)\right|}, 1)}$
                \EndIf
            \EndIf
            \State $D^{(1)}_m = 0$, $D^{(1)}_n = 0$, $D^{(2)}_m = 0$, and $D^{(2)}_n = 0$
        \Else
            \State $D^{(1)}_m \leftarrow D^{(1)}_m + \frac{\partial \mathcal{E}^{i,m}}{\partial e} $ \Comment{Accumulate first derivative in m-window}
            \State $D^{(1)}_n \leftarrow D^{(1)}_n + \frac{\partial \mathcal{E}^{i,n}}{\partial e}$ \Comment{Accumulate first derivative in n-window}
            \State $D^{(2)}_m \leftarrow D^{(2)}_m + \frac{\partial^2 \mathcal{E}^{i,m}}{\partial e^2} $ \Comment{Accumulate second derivative in m-window}
            \State $D^{(2)}_n \leftarrow D^{(2)}_n + \frac{\partial^2 \mathcal{E}^{i,n}}{\partial e^2} $ \Comment{Accumulate second derivative in n-window}
        \EndIf
        \State \textbf{return} $\mathcal{E}^{{i,m}^{\ast}}_e$
    \end{algorithmic}
    \end{algorithm}
    
\section{Service Fee Credits}
    In the work thus far, we presented how the PoEL protocol described in this work enables blockchains with intrinsic CDAs to attract, and efficiently and sustainably retain capital using budgeted financial incentives. We now explore how the underlying incentive structures used therein can be extended to incentivise the use of auxiliary services (Assuming that the network provides additional services beyond those related to CDA) embedded in the adopting blockchain network (e.g. smart contract transactions) linked to the new capital, with an additional focus on catalysing sustainable, and efficient usage of the network (incl. its resources).\\
    \\
    This is achieved by endeavouring to expand the user base, which not only enhances the network's decentralisation but also sustainable economic rationality\footnote{A higher number of users and transactions helps reduce the fixed cost per transaction, which accounts for gas fees and infrastructural expenses, and thus additionally bolsters the efficiency of the network's financial and infrastructural resources. This can be distributed over an increased number of transactions if the resources are productively utilised.}. Therefore, with more users and transactional volume, there is a better utilisation of liquidity submitted to the CDAs. Furthermore, it is clear that for the sustainable viability of a blockchain network, it must have a foundational number of users and transactions. However, this presents a circular dilemma: users and transactions gravitate towards networks already perceived as viable, meaning a network requires users to generate value, yet it needs to offer value to attract users.\\
    \\
    To solve this quandary, we introduce service fee credits which serve the dual purpose of dynamically attracting capital, and incentivising initial network usage, which we hope is extenuated with network effects - a phenomenon when a product or service subject to network effects becomes more valuable as more people use it, e.g. in a social network \cite{afuah2013network,majumdar1998network}. These credits complement, not replace, staking rewards, which are earned by LPs who use their assets to contribute to the adopting blockchain network's economic security.
    
    \begin{definition}[Service Fee Credits]
        Service fee credits are non-transferable, timed, and limited credits, which are denominated in the native staking token and linked to a user's wallet address, such that they can be used to offset future costs arising from the user's use of the network or be delegated to another user. These credits are granted to users when they submit capital to a CDA's liquidity pool(s) and PoS scheme.\\
        \\
        These service credits, represented by $D = \{\mathcal{V}, \mathcal{A},  \mathcal{C}^{ \mathcal{A}}_\mathcal{R} \}$, where $\mathcal{V} \in \mathop{\mathbb{I}}^{+}$ is the number of epochs for which credits are available for use in a period we call, rounds ($\mathcal{R} \in \mathop{\mathbb{I}}^{+}$), and $\mathcal{C}^{\mathcal{A}}_\mathcal{R} \in [0, \infty)$ represents replenishment caps, which represents the round-specific predetermined ceiling to which further credits can be added and used by the user address $\mathcal{A}$.
    \end{definition}

\begin{figure}[H]
        \centering
        \includegraphics[scale=0.6]{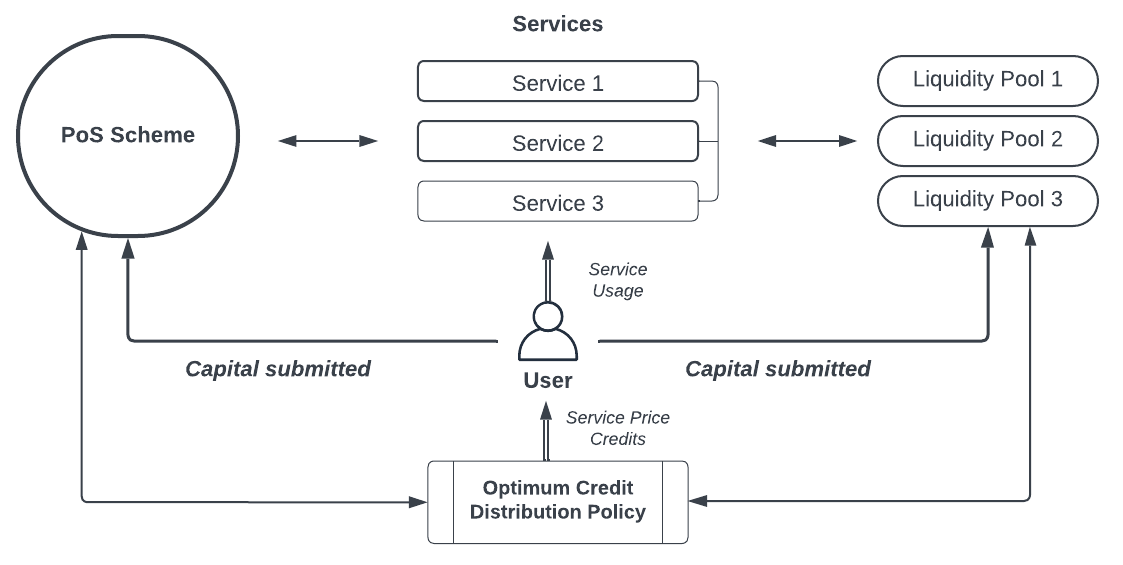}
        \caption{Service Based Capital attraction}
        \label{fig:sfc}
    \end{figure}

    \noindent
    For ease of reader's convenience, in Fig. \ref{fig:sfc}, we present schematics that exhibit how these credits fit in the context of the wider network, and how its distribution policy is closely linked with the incentive budget allocation process used in the PoEL protocol to attract capital.
    
    \noindent
    Now, in the definition of service fee credits, we introduced the notion of round-specific ($\mathcal{R}$) replenishment caps ($\mathcal{C}^{\tilde{X}^i}_{\mathcal{R}}$), which change with time and are linked to the nature of assets that comprise the capital submitted to the network. This is calculated as follows:
    \begin{equation}
        \mathcal{C}^{\tilde{X}^i}_\mathcal{R} = S_{\mathcal{R}}^{\tilde{X}^i} \cdot P^{\tilde{X}^i}_{\mathcal{R}} \cdot \gamma^{\tilde{X}^i}_R,
    \end{equation}
    \noindent
    where $\gamma^{\tilde{X}^i}_\mathcal{R}$ is the rate which dynamically determines the portion of the total submitted capital (in asset $\tilde{X}^i$) that can be used to replenish the capital provider's credit account at the beginning of a particular round; $S_{\mathcal{R}}^{\tilde{X}^i}$ is the size of the LP-token reserve at the beginning of a round; and $P^{\tilde{X}^i}_\mathcal{R}$ is the price of LP-token $\tilde{X}^i$ at the beginning of a round.\\
    \\
    As also stated in their definition, service fee credits are limited, meaning they are subject to a dynamic (periodic) limit ($\mathcal{C}^{\mathcal{A}}_\mathcal{R}$), which can be computed for a specific user as follows:
    
    \begin{equation}
        \mathcal{C}^{\mathcal{A}}_{\mathcal{R}} = \sum_{q=1}^n s^{{\mathcal{A}}^q}_\mathcal{R} \cdot \mathcal{C}^q_R,
    \end{equation}

    \noindent
    where $n$ is the number of LP token reserves in the system, and $s^{{\mathcal{A}}^q}_\mathcal{R}$ is the share of the user in the $q$-th asset's LP token reserve. Naturally, given some uncertainty in the user behaviour, we could use a stochastic differential equation to model the inherent uncertainty, e.g.: $d \mathcal{C}^{\mathcal{A}}_\mathcal{R} = \mu(\mathcal{C}^{\mathcal{A}}_\mathcal{R}, t)dt + \sigma(\mathcal{C}^{\mathcal{A}}_\mathcal{R},t)dW_t$, where $dW_t$ is a Wiener process term representing random fluctuations, and $\mu$ and $\sigma$ represent the drift and volatility terms, respectively. Furthermore, the system could incorporate a time-based locking incentive mechanism for users regarding replenishment cap allocation, akin to the framework suggested in Incentive 1, which will further encourage extended locking periods.\\
    \\
    Finally, we calculate the point-in-time ($t$) balance ($\mathcal{B}^{\mathcal{A}}_{t} \in [0, \infty)$) of the service fee credits for a particular account ($\mathcal{A}$) in a given round, as follows:
    \begin{equation}
        \mathcal{B}^{\mathcal{A}}_{t} = \mathcal{C}^{\mathcal{A}}_{\mathcal{R}} - \sum_{\mathcal{S}=1}^{Vs} N^{{\mathcal{D}}^S}_{{\mathcal{A}}_\mathcal{R}}.
    \end{equation}

    \noindent
    where $Vs$ is the number of network services for which service fee credits can be used; and $N^{{\mathcal{D}}^S}_{{\mathcal{A}}_\mathcal{R}}\in [0,\infty)$ is the amount of service fee credits that have been utilised by $\mathcal{A}$-th user address to pay for the service $\mathcal{S}$ in the round $\mathcal{R}$. \\
    \\
    Having defined the baseline concepts, we now consider the impact of bounding time when the service fee credits can be used. Therefore, on a ceteris paribus basis, the change (replenishment) in credit ($\Delta \mathcal{D}^{\mathcal{A}}_{\mathcal{R}}$) at the beginning of a round can be calculated as follows:
    \begin{equation}
        \Delta \mathcal{D}^{\mathcal{A}}_{\mathcal{R}+1} =  \mathcal{C}^{\mathcal{A}}_{\mathcal{R}} - \mathcal{B}^{\mathcal{A}}_{T'}
    \end{equation}
    
    \noindent
    where $T'$ is the last timestep of the round $\mathcal{R}$.\\
    \\
    Altogether, from a user's perspective, their risked capital is eligible for three streams of revenues - (i) PoS rewards, (ii) fees generated from CDAs, and (iii) service fee credits. However, since service fee credits are time-bound, it can be said that at the end of each round when their credits expire, unutilised credits are a source of cost, and therefore, a user's optimal return maximisation strategy ought to include full utilisation of awarded credits, which would aid the network in bootstrapping userbase and network activity.\\
    \\
    \textbf{Service Fee Credit Budgeting}\\
    \\
    The PoEL protocol presented a budgeted reward distribution policy, which was used to tactfully incentivise staked capital, which can be modified to budget any credits awarded.\\
    \\
    For rounds $\mathcal{R}> 1$, the total budget ($B_R \geq 0,s.t. B_{t=1} = 0$) available to finance such credits is:
    \begin{equation}
        B_\mathcal{R} = B_{\mathcal{R}-1} + \Delta{B}_\mathcal{R} - \sum^g_{i=1} \mathcal{C}^i_{\mathcal{R}-1},
    \end{equation}

    \noindent
    where $g$ is the highest index of LP tokens; and $\Delta{B}_\mathcal{R}$ is the addition to the credit budget in $\mathcal{R}$-th epoch, which can be determined using the state of the treasury and governance mechanisms.\\
    \\
    If the service fee credit program lasts for a defined number of rounds, and the budget is not static with time, then it can be stated (ex-post) that if the protocol is successful in attracting liquidity in earlier stages of its life, then $\frac{1}{\mathcal{R}} \int_{0}^{\mathcal{R}} f'(x) dx < 0$, meaning that in a mature state $\frac{\partial^2 \Delta B}{\partial \mathcal{R}^2} < 0$, implying that in earlier stages of the protocol, the protocol would seek to invest more in the incentive program and as the network matures, curtail any benefits. In other words, the additional amount ($\Delta B_\mathcal{R}$) needed to finance the incentive program would decrease at an increasing pace over time. This amount ($\Delta B_\mathcal{R}$) can be ascertained by inter-temporarily harmonising the network's fiscal objectives (a component of a blockchain network's fiscal policy), using a constrained optimisation problem which seeks to maximise the incentives distributed, under the constraint of network's survivability (defined in terms of the number of users associated with their auxiliary services and CDAs). In essence, this optimisation problem would enable us to ascertain the total (lifetime) distributable budget, and point-in-time budget, using the network's current state (e.g. including the number of users, the transactional volume, the number of services, etc.), and its precise functional form using the aforementioned state variables. Furthermore, using the framework presented in Sec. \ref{sec:dra}, we can ascertain the precise amount of credits available for distribution in each round, based on the distance of state variables to its aspired optimal value. Once a round-specific budget ($B_R$) is ascertained, the credit incentive mechanism can use the baseline PoEL protocol's variable lending fee modelling to ascertain the correct amount of incentives (by way of estimating the asset-specific replenishment cap) that can be provided for each unit of liquidity brought to the network.

\newpage
\section{List of Notations}

    \begin{longtable}{|c|p{10cm}|}    
    \caption{List of Notations} \label{table:notations} \\
    \hline \textbf{Notation} & \textbf{Description} \\ \hline 
    \endfirsthead
    
    \multicolumn{2}{c}
    {{\bfseries Table \thetable\ continued from previous page}} \\
    \hline \textbf{Notation} & \textbf{Description} \\ \hline 
    \endhead
    
    \hline \multicolumn{2}{|r|}{{Continued on next page}} \\ \hline
    \endfoot
    
    \endlastfoot
    
    \hline
    $t$ & A timestep, representing a specific point in time when a change in state variables is observed or input parameters are set. \\
    \hline
    $e$ & An epoch, representing a short yet relatively fixed period akin to block time representing instances when the protocol executes critical updates. \\
    \hline
    $e^l$ & Represents the number of epochs for which l-th LP's capital is locked.\\
    \hline
    $\epsilon$ & The maximum number of epochs for which the liquidity mining program is operational.\\
    \hline
    $e_{\text{mid}}$ & The mid-point value around which the transition from the minimum to maximum incentive starts to become significant.\\
    \hline
    $m, n$ & Distinct time windows for tracking changes in the moving average capital efficiency rate.\\
    \hline
    $V$ & Traded volume.\\
    \hline
    $\mathbf{S}$ & A vector of state variables specific to the protocol.\\
    \hline
    $P_i$ & The price of an asset at time t = i.\\
    \hline
    $Sp$ & Asset-specific bid-ask spread.\\
    \hline
    $V^{t_1:t_2}$ & Trading volume of an asset over a prespecified look-back window $[t_1,t_2]$.\\
    \hline
    $\mathbf{R}^i_{1:\epsilon}$ & A vector of total reward budget distributed to LPs in a particular CDA across epochs.\\
    \hline
    $\mathcal{L}_e^i$ & The (nominal) capital available in participating liquidity pools of the i-th CDA in the e-th epoch.\\
    \hline
    $\mathcal{I}_o^e$ & Incentives distributed for asset pool $o$ in $e$-th epoch.\\
    \hline
    $\mathcal{I}_t^{l, \max}$ & Maximum incentives available to distribute to a particular LP.\\
    \hline
    $\mathcal{I}_t^{l, \min}$ & Minimum incentives available to distribute to a particular LP.\\
    \hline
    $R_e^i$ & The optimal epoch-specific reward for LPs of i-th CDA.\\
    \hline
    $R_e^l$ & Represents the associated incentive.\\
    \hline
    $R_e^{l^{\max}}$ & Represents the maximum incentive available to distribute to a particular LP at a particular epoch.\\
    \hline
    $R_e^{l^{\max}}$ & Represents the minimum incentive (a portion of $R^{l^{\max}}_e$) available to distribute to a particular LP at a particular epoch.\\
    \hline
    $R^i_{{min}_e}$ & The minimum staking reward that needs to be distributed in the epoch $e$ as part of the liquidity bootstrapping program and is based on a system parameter.\\
    \hline
    $SR_e$ & Staking rewards earned.\\
    \hline
    $SRR_e$ & The staking reward rate for the adopting blockchain in the e-th epoch.\\
    \hline
    $\mathbf{I}_e$ & Vector of interest rates charged from rewards distributable for staked NST.\\
    \hline
    $DR^{{\tilde{X}}^i}_e$ & Rewards distributable to an LP's token reserves for an asset $\tilde{X}^i$ in an epoch.\\
    
    % Asset and Liquidity Specific Parameters
    \hline
    $S^{\tilde{X}^i}_{e_v}$ & LP Token reserve, representing the size of an LP's token reserve in a particular epoch, representing a unique collateral pool mapped to the $v$-th validator’s address.\\
    \hline
    $P^{\tilde{X}^i}_e$ & The price of LP asset $\tilde{X}^i$ denominated in the native staking token at the epoch $e$.\\
    \hline
    $L^{\tilde{X}^i}_e$ & The loaned amount of the native staking token, using collateral in asset $\tilde{X}^i$.\\
    \hline
    $L_e$ & The total amount of native staking tokens loaned to borrowers.\\
    \hline
    $Q$ & Size of buy (or sell) order.\\
    \hline
    $IP^{\tilde{X}^i}_e$ & Nominal interest payable.\\
    \hline
    $\mathcal{T}_e$ & The amount of native staking tokens which can be borrowed in any epoch.\\
    \hline
    $\rho^{X^i_{LP}}_e$ & Collateralisation rate, quantifying the amount of collateral an LP needs to deposit to be able to borrow desired native staking tokens.\\
    \hline
    $\rho_{min}$ & A system parameter denoting the minimum collateralisation rate for all pools.\\
    \hline
    $w^{\prime}$ & Weights associated with each conceptual objective.\\
    \hline
    ${w^{{\tilde{X}}^i}_e}$ & The weight of the asset in the basket in e-th epoch.\\
    \hline
    $w^{\tilde{X}^{i^{\ast}}}_t$ & target weights of asset $\tilde{X}^i \neq 0$.\\
    \hline
    $\lfloor{\mathcal{W}}\rfloor, \lceil{\mathcal{W}}\rceil$ & Minimum and maximum permissible weights of an asset.\\
    \hline
    $\mathcal{W}^{\tilde{X}^i}_e$ & Weight associated with the capital efficiency of asset $\tilde{X}^i$ in the portfolio.\\
    
    % Risk and Market Parameters
    \hline
    $\sigma$ & Realised standard deviation of a particular asset.\\
    \hline
    $\lambda_{{OL}_1}$& Probabilistic threshold for the likelihood of the occurrence of prespecified operational risk events.\\
    \hline
    $\phi_{j}$ & Market impact function of j-th asset.\\
    \hline
    $MI^{\phi_{NST}, V}_{NST}$ & market impact of executing a fixed notional value $V$ in the native staking token with a specific market impact function.\\
    \hline
    $\eta^{\tilde{X}^i}_e$ & The risk scaling factor for asset $\tilde{X}^i$.\\
    \hline
    $b$ & Steepness of collateral diversification curve.\\
    \hline
    $\zeta$ & The smallest portion of staking rewards that need to be distributed to bootstrap the liquidity of CDA.\\
    \hline
    $\mathcal{E}^{{i,m}^\ast}_e$ &Target (moving average) capital efficiency rate to enhance a CDA's average capital efficiency.\\
    \hline
    $\mathcal{E}_0^{{i,m}^\ast}$ & Initial target capital efficiency rate.\\
    \hline
    $\frac{\partial \mathcal{E}^{i,m}}{\partial e}, \frac{\partial^2 \mathcal{E}^{i,m}}{\partial e^2}$ & First and second derivatives of the moving average of capital efficiency rate over the last epoch, respectively.\\
    \hline
    $\Upsilon$ & Global parameter regulating the speed of negative adjustments.\\
    \hline
    $\psi$ & Global parameter regulating the speed of positive adjustments.\\
    
    % Operational and Structural Parameters
    \hline
    $r_t$ & Realised return of a particular asset.\\
    \hline
    $\kappa$ & Scaling factor adjusting the steepness of the sigmoid function controlling the incentive awarded based on tenure, in comparison to the target tenure.\\
    \hline
    $\varrho$ & Parameter adjusting the steepness of available incentive's relationship with the tenure of deposit (i.e. locking time).\\
    \hline
    $k$ & The scaling factor adjusting steepness of the sigmoid curve governing the reward in relation to the locking time.\\
    \hline
    $\nu$ & A new parameter that adjusts the curve's shape, providing additional control over how rapidly the incentive increases as the locking time approaches and surpasses mid-point.\\
    \hline
    $E^i_{1:\epsilon}$ & The capital efficiency of i-th CDA.\\
    \hline
    $RP_e$ & The size of reward pool.\\
    \hline
    $\mathcal{L}^{i^X}_e, \mathcal{L}^{i^X}_{{LP}_e}$ & Liquidity available in asset pool $X$ of the CDA $i$ at the epoch $e$.\\
    \hline
    $\mathcal{E}^{{i,m}^X}_{e}$ & Capital efficiency rate over a period of time (look-back window of length - $m$ epochs).\\
    \hline
    $\chi$ & The sensitivity to weight deviations in the collateral basket from its target.\\
    \hline
    $\mathcal{K}$ & System parameter governing the convexity of the relationship between weights and the capital efficiency rate.\\
    \hline
    $\mathcal{G}$ & Factor governing the desirability of a higher capital efficiency asset in the portfolio.\\
    \hline
    $I^{\tilde{X}^i}_e$ & Interest rate charged for each asset $\tilde{X}^i$.\\
    \hline
    $R_e$ & Reward budget for the $e$-th epoch.\\
    \hline
    $g$ & Highest index of LP tokens from different pools of the $i$-th CDA.\\
    \hline
    $R^j_e$ & Staking rewards distributable to LP token reserves for a staked NST borrowed using the $j$-th asset.\\
    \hline
    $E^{i^X}_e$ & Capital efficiency rate for an asset pool within a CDA.\\
    \hline
    \end{longtable}

\bibliography{main.bib}
\bibliographystyle{plain}

\end{document}